\documentclass{elsarticle}
\usepackage{braket}
\usepackage{amssymb}
\usepackage{graphicx}
\usepackage{amsmath}
\usepackage{color}
\usepackage{hyperref}

\begin{document}
\title{Study of entanglement in Ne, Mg, and Si isotopic chains}

\author{Rohit M. Shinde\footnote{Email: rohit$\_$ms@ph.iitr.ac.in} and Praveen C. Srivastava\footnote{Email: praveen.srivastava@ph.iitr.ac.in}}
\address{Department of Physics, Indian Institute of Technology Roorkee, Roorkee Uttarakhand - 247667, India}


\date{\today}
\begin{abstract}
In this article, we have presented the results for mode entanglement and its evolution in Ne, Mg, and Si isotopic chains. A detailed study has also been conducted on the role of nuclear forces with the assistance of the spin-tensor decomposition of two-body interaction into three components: central, spin-orbit, and tensor. This study can shed light on the high entanglement entropy observed for $N=Z$ nuclei and the importance of the tensor component on the proton-neutron entanglement entropy. 
\end{abstract}
\maketitle

\section{Introduction}
The atomic nucleus is a self-bound system of protons and neutrons, making it an ideal candidate for studying quantum many-body phenomena. Configuration-interaction methods are well-suited for solving such nuclear many-body problems \cite{Caurier_2005}. As the number of nucleons increases, the complexity increases, giving rise to collectivity and other intricate relationships between different configurations of nucleons within a nucleus. Recent advancements in quantum information science have offered new insight into the role of entanglement in studying nuclear systems \cite{Tichai_2023,Carl_2024,Kruppa_2022}. Quantum computing, with its inherently quantum nature, presents a promising framework for simulating complex nuclear dynamics. Algorithms designed for quantum processors are particularly well-suited for capturing entanglement and correlations in many-body systems \cite{Chandan_2023,Denis_2024,Perez2_2023}. Entanglement is a measure of dependence between different systems or different partitions within the same system; greater dependence implies higher entanglement between those systems. Such correlations can be described with the help of entanglement entropy \cite{CJ_2023}.
 
Different partitions can be applied to the wavefunction of atomic nuclei to provide details about the type of entanglement that exists between the nucleus's distinct components \cite{Carl_2021}. The atomic nucleus provides a natural bi-partition where protons and neutrons can be considered as two different sub-systems. The many-body basis $\ket{\psi}$ is written as a linear combination of the Slater determinant basis $\{\ket{\alpha}\}$, which is further decomposed into the neutron ($\nu$) and proton ($\pi$) components \cite{Johnson_2025}:

\begin{equation}
 \ket{\alpha} = \ket{\phi_{\pi}} \otimes \ket{\phi_{\nu}} 
\end{equation}
    
\begin{equation}
    \ket{\psi} = \sum_{\pi,\nu}c_{\pi,\nu}|\phi_{\pi} \rangle \otimes |\phi_{\nu} \rangle. 
\end{equation}
In the present work, to calculate the entanglement entropy, we have utilized the von Neumann entropy  \cite{CJ_2024,chuang} $S$ = -tr $\rho^{red} \log\rho^{red}$ of a few selected $sd$-shell nuclei and their isotopic chains using USDB interaction \cite{Brown_2006}. All the calculations to obtain the coefficients of the wavefunctions were performed using BIGSTICK, a configuration-interaction shell model code \cite{Johnson_2018}. 

In section \ref{Mode_Ent}, we have introduced the formalism for mode entanglement entropy and its evolution. In section \ref{spin-tensor}, we have performed the spin-tensor decomposition of the USDB  interaction, and in section  \ref{results}, we have calculated the proton-neutron entanglement entropy for the corresponding interactions.

\section{Mode entanglement entropy in $sd$ shell nuclei}
\label{Mode_Ent}

Mode entanglement entropy is a measure in quantum many-body systems to quantify entanglement between different modes, typically single-particle states rather than particles \cite{Perez_2023, Kruppa_2021}. This distinction is necessary in systems containing indistinguishable particles, such as protons and neutrons in atomic nuclei. The reason is that, for indistinguishable particles, the many-body Hilbert space lacks a simple tensor-product structure, which makes it difficult to partition the system. This issue is addressed using the Fock space representation, where states are described in terms of occupation numbers. In this framework, entanglement is defined between single-particle states (or modes), not individual particles \cite{Carl_2021}.

For the nuclear systems, the entanglement of a mode with the rest of the modes is evaluated with the help of the mode-reduced density matrix $\rho$ as

\begin{equation}
    \rho_i = \begin{pmatrix}
             1-\gamma_{ii} & 0 \\
             0 & \gamma_{ii}
            \end{pmatrix},
\end{equation}
where $\gamma_{ii} = \bra{\psi}a_i^{\dagger}a_i\ket{\psi}$ is the occupation probability for the corresponding single-particle state $i$. The off-diagonal elements vanish due to particle number conservation. The von Neumann entropy characterizing the entanglement between a single-particle state and other states is given by

\begin{equation}
    S_i = - (1 - \gamma_{ii}) \ln(1 - \gamma_{ii}) - \gamma_{ii} \ln(\gamma_{ii}).
\end{equation}
Single-orbital entropy attains a maximum value of ln(2) when that orbital has an occupation probability of 0.5, indicating that the orbital is maximally entangled with the rest of the system. When the orbital is fully occupied ($\gamma_{ii}=1$) or empty ($\gamma_{ii}=0$), the entropy vanishes, suggesting low correlations with the rest of the orbitals. \par
\begin{figure}[!htbp]
    \centering
    \includegraphics[width=1.00\textwidth]{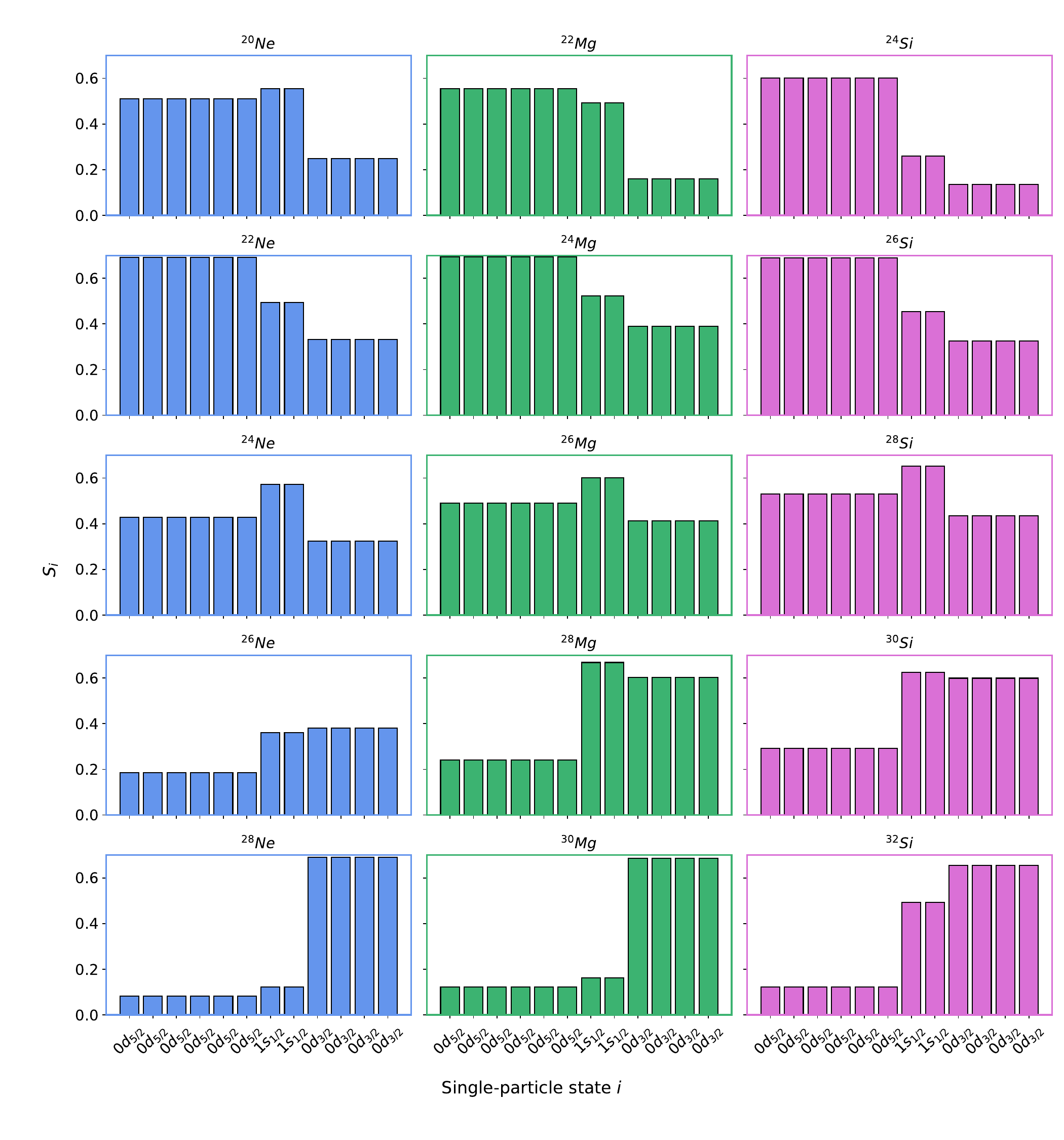}
    \caption{Single-orbital entanglement entropy $S_i$  for Ne, Mg, Si isotopic chains, where $i=(n_i,l_i,j_i,m_i,\tau_i=-1/2)$, these correspond only to single-neutron states of the isotopes.}
    \label{fig:single_particle_ent}
\end{figure}

We have shown in Figure \ref{fig:single_particle_ent} the entanglement entropies $S_i$ of the $sd$ shell single-particle orbitals for the isotopic chains of Ne($^{20-28}$Ne), Mg($^{22-30}$Mg), and Si($^{24-32}$Si) in the ground states. As the shell model Hamiltonian follows spherical symmetry, the entanglement entropy corresponding to the same $nlj$ is always equal. The calculations are limited to neutron orbitals, as we aimed to show the trend in single-orbital entropy with increasing neutron numbers along an isotopic chain. The entropy for the \( d_{5/2} \) orbital peaks at \( N = 12 \) in all isotopes since its occupation is close to half. For the \( s_{1/2} \) orbital, the maximum entropy is seen in \(^{24}\text{Ne} \), \(^{28}\text{Mg} \), and \(^{28}\text{Si} \). As more neutrons are added, the lower orbitals become fully occupied, which leads to a drop in their single-orbital entropy. The \( d_{3/2} \) orbital reaches its highest entropy at \( N = 18 \), again due to the occupation being close to 0.5.

\section{Role of spin-tensor decomposition on entanglement entropy}
\label{spin-tensor}
To understand how different parts of the nuclear force affect the evolution of proton–neutron entanglement entropy, we have followed the approach outlined in Ref. \cite{Smirnova_2012} for the spin–tensor decomposition. Within this framework, we performed a spin–tensor decomposition of the USDB interaction. Any two-body interaction can be decomposed into scalar, vector, and tensor components:

\begin{equation}
    V = \sum_{k=0,1,2}(S^{(k)}.Q^{(k)})=\sum_{k=0,1,2}V^{(k)},
\end{equation}
where $S^{(k)}$ are spherical tensors of rank $k$ in spin space and $Q^{(k)}$ are spherical tensor of rank $k$ in coordinate space. The term with $k=0,1,$ and $2$ corresponds to the central, vector (spin-orbit), and tensor components of the matrix elements, respectively. To perform decomposition, the two-body matrix elements are first converted from the $jj$-coupling scheme to the $LS$-coupling scheme as follows

\begin{align}
\langle n_a l_a n_b l_b : LS, JT | V | n_c l_c n_d l_d : L'S', JT \rangle 
= \sum_{j_a j_b} \frac{N_{l_a l_b}}{N_{j_a j_b}} \sqrt{(2L + 1)(2S + 1)(2j_a + 1)} \nonumber \\
\times  \sqrt{2j_b + 1}
\left\{
\begin{array}{ccc}
l_a & l_b & L \\
\frac{1}{2} & \frac{1}{2} & S \\
j_a & j_b & J
\end{array}
\right\}
\sum_{j_c j_d} \frac{N_{l_c l_d}}{N_{j_c j_d}} \sqrt{(2L' + 1)(2S' + 1)(2j_c + 1)(2j_d + 1)} \nonumber \\
\times 
\left\{
\begin{array}{ccc}
l_c & l_d & L' \\
\frac{1}{2} & \frac{1}{2} & S' \\
j_c & j_d & J
\end{array}
\right\}
\langle n_a l_a j_a n_b l_b j_b : JT | V | n_c l_c j_c n_d l_d j_d : JT \rangle,
\end{align}

where, normalization factor $N_{l_al_b}$ is

\begin{equation}
N_{l_a l_b} = \frac{1}{\sqrt{2(1 + \delta_{l_a l_b})}}.
\end{equation}
Using the LS-coupling scheme, the matrix elements of each component of the two-body interactions are written as 
\begin{align}
\langle n_a l_a n_b l_b : LS, J'T | V^{(k)} | n_c l_c n_d l_d : L'S', J'T \rangle 
= (2k + 1)(-1)^{J'}
\left\{
\begin{array}{ccc}
L & S & J' \\
S' & L' & k \\
\end{array}
\right\}
\nonumber \\
\times \sum_J (-1)^J (2J + 1)
\left\{
\begin{array}{ccc}
L & S & J \\
S' & L' & k \\
\end{array}
\right\}
\langle n_a l_a n_b l_b : LS, JT | V | n_c l_c n_d l_d : L'S', JT \rangle.
\end{align}
Next, we examined how different components of the interaction contribute to and influence the evolution of ESPEs, which, in turn, drive the evolution of proton–neutron entanglement entropy.

The single-particle energies (SPE) are influenced by the contribution of nucleons outside the inert core, which is due to the mean-field effects created by these valence nucleons. This helps us indicate the positions of the single-particle orbitals that play a key role in the evolution of the shell gap \cite{Otsuka_2020}. In this work, we have shown how the proton effective single-particle energies (ESPEs) for Si isotopes, based on the USDB interaction, evolve in relation to the neutron ESPEs. The proton ESPEs are given by 

\begin{align}
\epsilon(j_p) &= \epsilon^0(j_p) + \sum_{j_n} V_{pn}^m(j_p, j_n) n_{j_n}, 
\end{align}

\noindent where $n_{j_n}$ are neutrons occupying the orbitals $j_n$ and  \( V_{pn}^m(j_p, j_n) \) is the monopole matrix element between \( j_p \) and \( j_n \), which is given by

\begin{align}
V_{pn}^m(j_p, j_n) 
&= \frac{\sum\limits_{J = |j_p - j_n|}^{j_p + j_n} (2J + 1)V_J(j_p j_n j_p j_n)}{\sum\limits_{J = |j_p - j_n|}^{j_p + j_n} (2J + 1)}. 
\end{align}
In the next section, we will discuss the effect of spin-tensor decomposition and the evolution of ESPE on proton-neutron entanglement entropy.

\section{Results}
\label{results}
\begin{figure}[!htbp]
 \centering
  \begin{tabular}{cc}
     \includegraphics[width=0.58\textwidth]{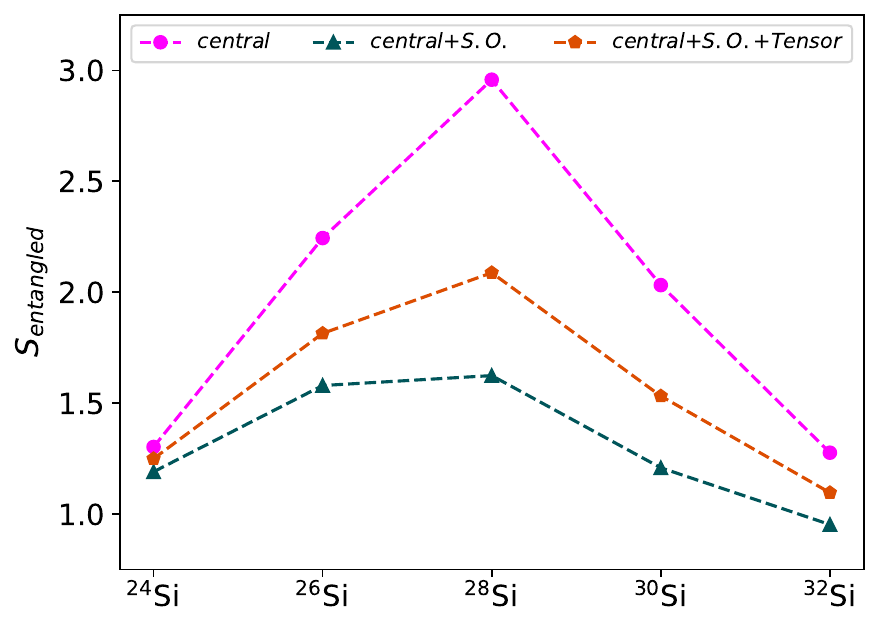}
  \end{tabular}
  \caption{Entanglement entropies of Si isotopic chain. The comparisons are done between the contributions from central, spin-orbit, and tensor components of the USDB interaction.}
  \label{fig:Si_usdb_decomp}
\end{figure}

First, we have performed proton-neutron entanglement calculations in Si isotopes after decomposing the USDB interaction into central, spin-orbit, and tensor components. All the calculations for the entanglement are performed in the M-scheme. 
The ground state entanglement entropies for $^{24\text{--}32}$Si are presented in Figure~\ref{fig:Si_usdb_decomp}.

\begin{figure}[!t]
  \begin{tabular}{cc}
    \includegraphics[width=0.48\textwidth]{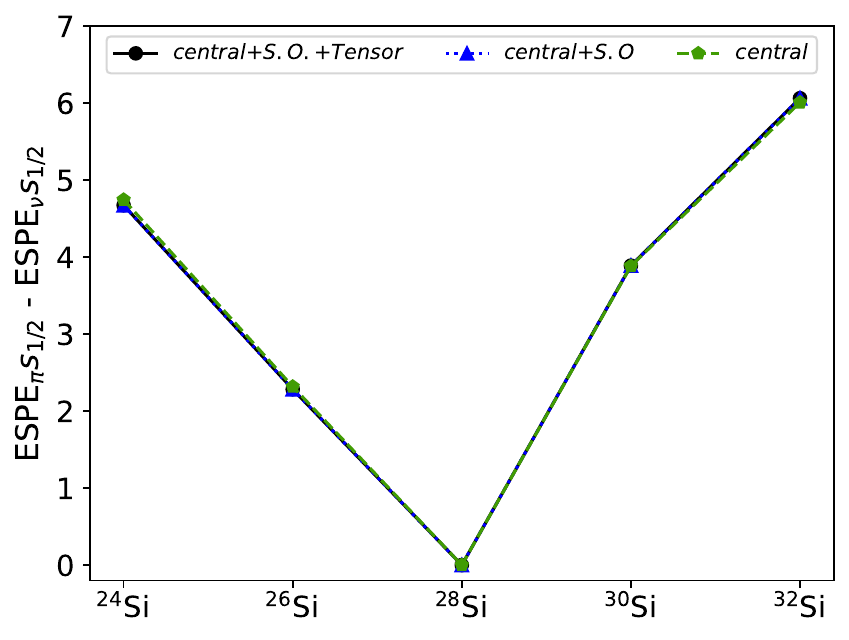} 
     \includegraphics[width=0.48\textwidth]{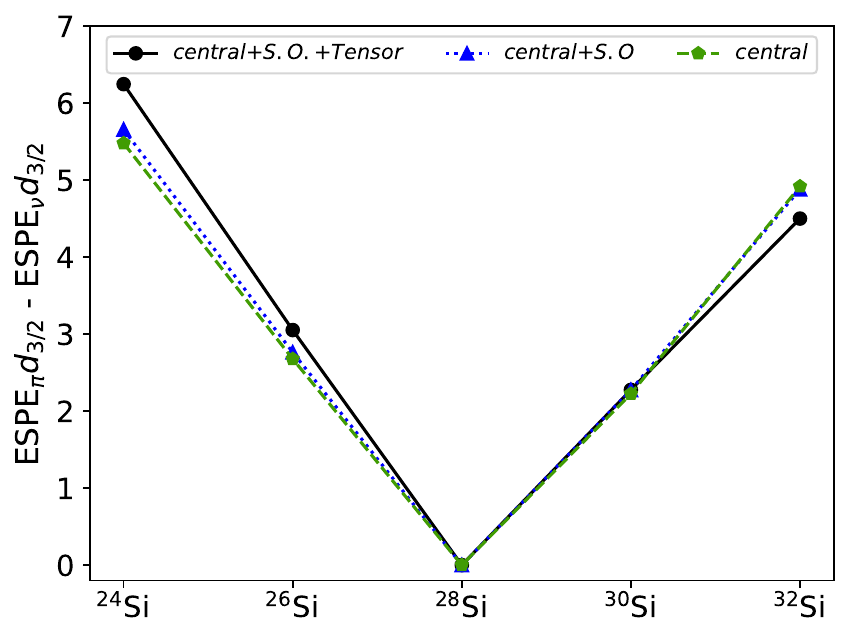}\\
     \includegraphics[width=0.48\textwidth]{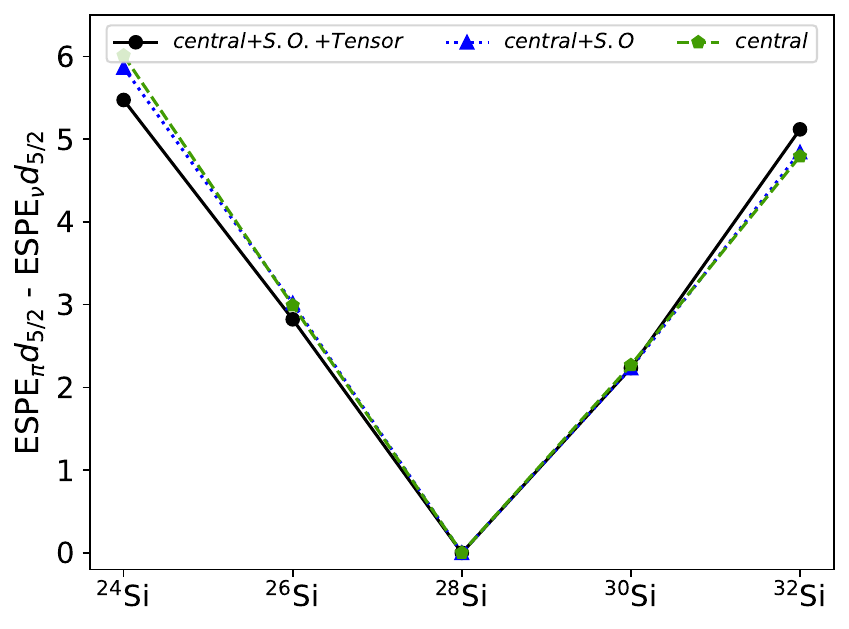} 
  \end{tabular}
  \caption{Difference between the proton effective single particle energies and neutron effective single particle energies for Si isotopes. The difference is taken between like-orbitals, and the comparison is between the spin-tensor decomposed parts of the USDB interaction.}
  \label{fig:espe_like_states}
\end{figure}

\begin{figure}[!htbp]
  \centering
     \includegraphics[width=0.50\textwidth]{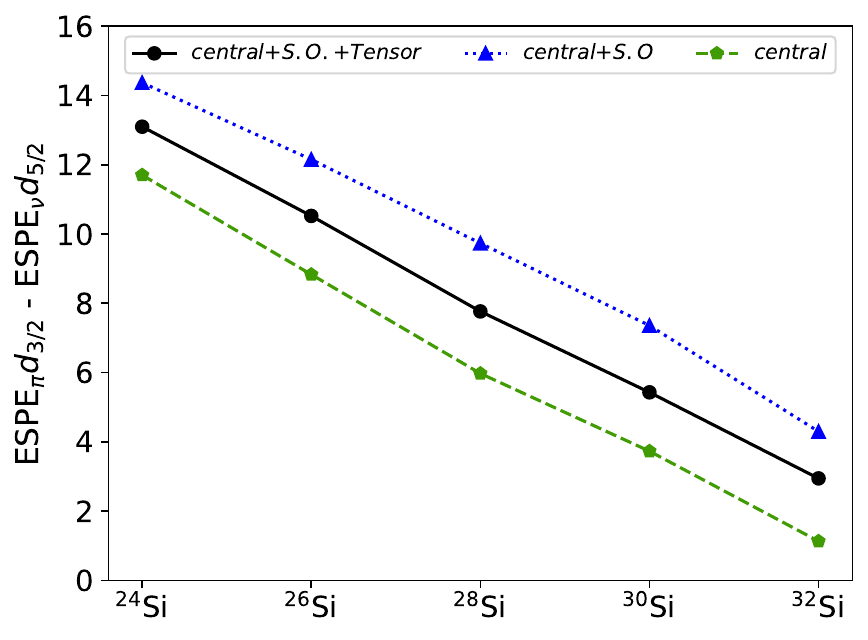}
  \caption{Difference between the effective single particle energies of proton and neutron is shown corresponding to $d_{5/2}$ and $d_{3/2}$ orbitals. The comparison is done between the spin-tensor decomposed parts of the USDB interaction.} 
  \label{fig:espe_tensor}
\end{figure}

The entanglement entropy reaches a maximum at $^{28}$Si for all three decomposed parts of the interaction, corresponding to the $N=Z$ case, and decreases symmetrically with a 
change in neutron number on either side. The overall trend among the three interactions exhibits a similar pattern. The central component yields the largest entanglement entropy across the isotopic chain, but when the spin-orbit term is introduced, the entropy drops significantly, and further inclusion of the tensor component causes the entropy to rise again. This trend can be explained by the energy differences between proton and neutron ESPEs along the isotopic chain across different orbitals. In Figure~\ref{fig:espe_like_states}, the energy difference between like-orbitals is calculated. It is observed that the difference between proton and neutron ESPEs tends toward zero at $N=Z$ ($^{28}$Si) for all three interactions, and increases as the neutron number deviates from this symmetry. This can be directly correlated with the increase in entanglement entropy when the energy difference is lower, which occurs at the $N=Z$ nuclei. To explain the reduction in entanglement entropy when the spin-orbit term is added, we calculate the difference between proton and neutron ESPEs for the $d_{5/2}$ and $d_{3/2}$ orbitals, as shown in Figure~\ref{fig:espe_tensor}. As seen in this figure, the spin-orbit term leads to the largest energy splitting across the entire isotopic chain, which reduces the entanglement entropy. The addition of the tensor component to the interaction attracts the $d_{5/2}$ and $d_{3/2}$ orbitals, increasing the entanglement entropy once again \cite{Otsuka_2005}. We do not consider the $s_{1/2}$ orbitals since they are unaffected by tensor forces, and therefore, the corresponding energy difference is not significant.

\begin{figure}[!htbp]
 \centering
  \begin{tabular}{cc}
      \includegraphics[width=0.58\textwidth]{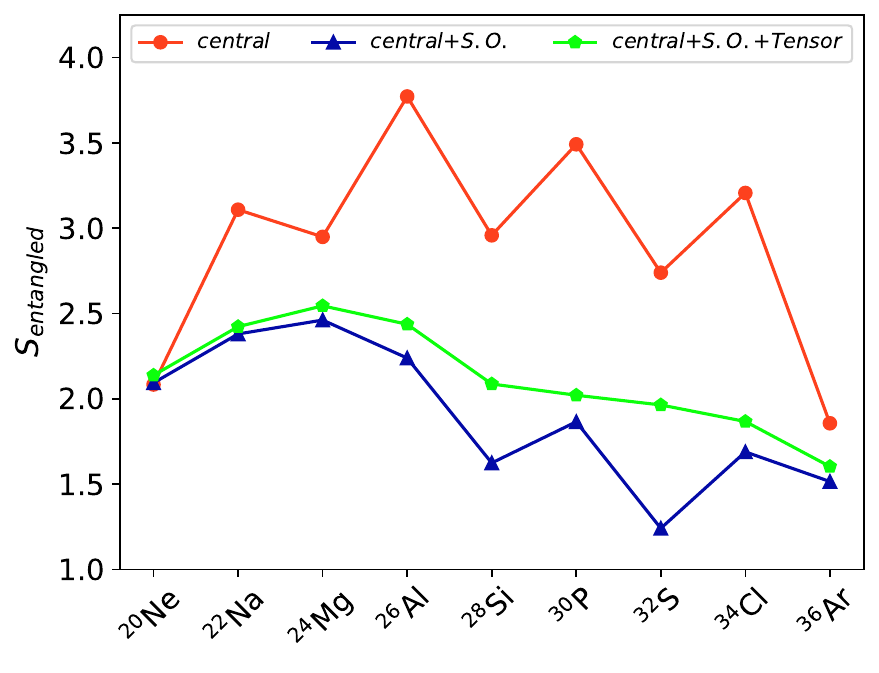}
  \end{tabular}
  \caption{Entanglement entropies for $N=Z$ nuclei in the $sd$-shell. The comparisons are done between the contributions from central, spin-orbit, and tensor components of the USDB interaction.}
  \label{fig:Si_usdb_decomp2}
\end{figure}

The ground state entanglement entropies for $N=Z$ nuclei in the $sd$-shell are also shown in Figure~\ref{fig:Si_usdb_decomp2}. The central component of the interaction exhibits an odd-even staggering, which can be attributed to the interaction between unpaired protons and neutrons in the odd-odd nuclei. This odd-even effect disappears
after the addition of the spin-orbit term, and a drop in entanglement entropy can be observed at the shell closures at $^{28}$Si and $^{32}$S. The inclusion of the tensor component significantly increases the entanglement entropy for $^{28}$Si and $^{32}$S, this is likely due to the attractive force between the $\pi (d_{5/2})$ and $\nu (d_{3/2})$ orbitals. 

\section{Conclusion}
In the present work, we have reported the shell model results of proton-neutron entanglement for $sd$ shell nuclei.
Our results have demonstrated that entanglement entropy reaches a maximum for the $N=Z$ condition, where the proton-neutron structure is most symmetric. This highlights the dependence of entanglement entropy on the effective single-particle energies of protons and neutrons. Decomposing the USDB interaction into central, spin-orbit, and tensor components reveals the underlying structural changes in the single-particle energies, which determine the evolution of entanglement entropy.

\section*{Acknowledgements}
We acknowledge financial support from MHRD, the Government of India. We would like to thank the National Supercomputing Mission (NSM) for providing computing resources of ‘PARAM Ganga’ at the Indian Institute of Technology Roorkee, implemented by C-DAC and supported by MeitY and DST, Government of India.



\end{document}